\documentstyle[12pt,aasms4]{article}

\slugcomment{Submitted to the Astrophysical Journal Letters.}

\lefthead{Richer {\it et al.}}
\righthead{The White Dwarf Cooling Age of M67}

\begin{document}

\title{The White Dwarf Cooling Age of M67}

\author{Harvey B. Richer\altaffilmark{1}, Gregory G. Fahlman\altaffilmark{1},
Joanne Rosvick and Rodrigo Ibata\altaffilmark{2}}
\affil{Department of Physics and Astronomy, University of British Columbia\\
 129-2219 Main Mall, Vancouver, BC, Canada V6T 1Z4}


\altaffiltext{1}{Visiting Astronomer, Canada-France-Hawaii Telescope, which
is operated by CNRS of France, NRC of Canada and the University of Hawaii.} 
\altaffiltext{2}{present address: European Southern Observatory, Garching, Germany.}


\begin{abstract}
A deep imaging survey covering the entire $23\arcmin$ diameter of the old open cluster M67 to $V = 25$ has been carried out using the mosaic imager (UHCam) on
the Canada-France-Hawaii Telescope. The cluster color-magnitude diagram (CMD)
can be traced from stars
on its giant branch at $M_{V} = +1$ down through main sequence stars at
least as faint as $M_{V} = 13.5$. Stars this low in luminosity have masses
below $0.15
M_{\odot}$. A modest white dwarf (WD) cooling sequence is also 
observed commencing
slightly fainter than $M_V = 10$ and, after correction for background galaxy 
and stellar field contamination, terminating near $M_V = 14.6$. The observed WDs follow quite closely a theoretical cooling sequence for $0.7 M_{\odot}$
pure carbon core 
WDs with hydrogen-rich atmospheres (DA WDs).
The cooling time to an $M_V$ of 14.6 for such WDs is $4.3$
Gyr which
we take as the WD cooling age of the cluster. A fit 
of a set of isochrones to the cluster CMD
indicates a turnoff age of $4.0$ Gyr.
The excellent agreement between these results suggests that ages
derived from white dwarf cooling
should be considered as reliable as those from
other dating techniques. The WDs currently contribute about 9\% of the total cluster mass but the number seen
appears to be somewhat low when compared with the number of giants observed 
in the cluster.
 
\end{abstract}




%

\section{Introduction}

The ages of stellar clusters are usually determined by fitting a set 
of theoretical isochrones to its CMD. When the stellar system 
is a globular cluster the age derived is usually considered an estimate
of the age of the Universe. Another estimate of the age of the
Universe, the expansion age, comes from the Hubble Constant 
with dependence on the density
parameter $\Omega$ and the cosmological constant $\Lambda$.  
These ages, 
derived by entirely independent techniques, generally do not
agree particularly well. The expansion age of a flat
Universe ($\Omega = 1$, $\Lambda = 0$) with $H_0 = 70$ is a relatively 
young 9.3 Gyr
whereas one globular cluster, M92, seems to be 14 Gyr old (\cite{pon98}). There
are some counter-claims that other globular clusters might be younger (\cite{cha98,gra98,rei98}) based on revised Hipparcos distances for
subdwarfs, but these are largely statistical studies and
the \cite{pon98} study did incorporate the new Hipparcos distances.

The above discussion illustrates that the age of the Universe, a critical
parameter in choosing amongst possible cosmological models, is still quite
uncertain. One approach to providing more confidence in an age estimate
is to determine it from a number of different techniques. The age of the
Universe could be estimated from the very old WDs found in
globular clusters in a manner similar to the age determination of 
the disk of the Galaxy (\cite{win86}), a technique
quite independent of isochrone fitting to cluster main sequences.
We have carried out some tests of the accuracy of WD cooling models in 
our investigation
of the WD population in the globular cluster M4 (\cite{ric95}, 1997). 
It now remains to obtain imaging data deep enough
to see the termination point of the WD cooling sequence in this and other
Galactic globular clusters. 

An additional test of the accuracy of WD derived ages 
is to obtain the age of an open cluster from its WDs and compare this
with the cluster turnoff age (see \cite{hip95} for an early effort in this direction). Many of the difficulties in determining
globular cluster ages are absent when considering open clusters. There is
empirical knowledge of the mass-luminosity relation for population I stars
and, unlike globular clusters, metallicity effects are generally not an
issue. The isochrones are
thus better constrained and main sequence fits of these models to real cluster
CMDs are more reliable.  
What is required here is a rich open cluster that
is nearby, relatively old (so that there is a substantial population of
WDs), at high Galactic latitude (to reduce field contamination) and lightly
reddened. The outstanding candidate is M67.

Studies of the CMD of M67 trace the history of such work from early
photoelectric observations done on a star-by-star basis (\cite{joh55,egg64}),
to deep photographic studies (\cite{rac71}), through to modern investigations
with CCDs (\cite{mon93}). Racine commented on the lack of WDs in his CMD 
(none were found in excess of the expected background) and
made the suggestion that none existed in the cluster brighter than
$M_V = 12.2$. This was in contrast to an earlier statement by \cite{baa63} that
about 3 dozen WDs existed in the cluster between magnitudes 20 and
22.7. The Montgomery et al. study found 5 objects bluer than $(B - V) = 0.0$ 
but these stars do not appear to lie along any recognizable cooling sequence.
The isochrone age
of M67 has been estimated by numerous authors with the range extending from
just older than 3 Gyr to almost 6 Gyr with the
most recent ages, incorporating the best stellar models, giving ages near 4 Gyr
(($m - M_V) = 9.59$, $E(B-V) = 0.05$, \cite{mon93}).
The total cluster mass has been estimated to be  
about $800 M_{\odot}$ (\cite{fra89,mat85,mon93}) and \cite{mat85}
indicated that up to half the mass of the cluster could be composed of WDs.
We comment on these estimates at the end of this contribution.  

\section{Observations and Reductions}

Most of the observations for the present work were obtained during
an excellent 5 night
run in January 1997 using the UH8k imager on the Canada-France-Hawaii Telescope.
This mosaic imager has 8000$\times$8000 15$\mu$ pixels which project to an area
of 28$\arcmin$$\times$28$\arcmin$ on the sky, just slightly larger 
than the 
23$\arcmin$ diameter of M67. 
Aside from a number of short exposures made to
avoid saturating the brightest stars in the cluster, all the images obtained
had an exposure time of 1200 sec. In total, 15 $V$ and 11 $I$ frames were obtained
centered on the cluster and 14 $V$ and 13 $I$ exposures were secured of a blank
field at the same Galactic latitude as M67 and about 1.2 degrees away 
in longitude. Numerous photoelectric standards (\cite{lan83}, 1992) were also
observed on all the 8 individual CCD chips making up the mosaic and independent
calibrations were derived for each chip. The photometric solutions had an
average scatter of 0.014 magnitudes in $V$ and 0.017 in $I$.
The blank field images were dithered so
that a median filtered image of them provided the flat field. The seeing was
generally quite good during the run becoming as excellent as 0$\farcs$5 and
degrading to worse than 1$\farcs$2 at times. 

The data were reduced using ALLFRAME (\cite{ste87}, 1994) with a quadratically
varying point spread function (PSF). The errors
in the photometry in the M67 field for stars in the magnitude bin 
$24 < V < 25$ were typically 0.06 magnitudes and for 
$25 < V < 26$ the 
errors were 0.10 magnitudes. In the blank field the equivalent numbers 
were 0.07 and 0.12 magnitudes.  

\section{The M67 Color-Magnitude Diagram}

In Figure 1 we present the $M_V$, $(V - I)_{\circ}$ color-magnitude diagram (CMD) of M67 
(left-hand panel) and the blank field (right-hand panel). 
The objects
plotted in both these diagrams only include images that appeared to be
stellar in appearance (see $\S$4.2). In the M67
CMD a distinct main sequence is observed beginning at the cluster turnoff
near $V = 12.6$ ($M_V = 3$) and extending to at least $V = 23.1$ ($M_V = 13.5$). The main sequence may extend to the limit of the data but the
paucity of stars at faint magnitudes on the main sequence make this 
difficult to determine. The turnoff is sharply defined and a subgiant
branch and red clump is observed, these latter stars taken from the compilation
of \cite{mon93} as they were saturated on our images. An extensive sequence is seen lying above
the M67 main sequence and there is little doubt that this is due to binaries
as has been noted by numerous other authors (e.g \cite{mon93}).
A few cluster blue stragglers are noted near $M_V = 1.5$, 
$(V - I)_{\circ} = 0.2$. 
The wall of stars lying below the M67 main sequence is composed of faint
field stars, most likely in the halo. The color of the bluest of these
field stars is redder than the M67 turnoff so these stars are almost certainly
older than M67. It is unlikely that their redder color comes from enhanced
metallicity with respect to M67 as the cluster has a metallicity close to 
solar. One final feature of the M67 CMD is a population of faint blue
stars that are stretched out along a sequence beginning near $M_V = 10.5$ and 
projecting into the smear of objects (faint unresolved
galaxies) centered at $V = 24$ and $(V - I)_{\circ} = 1$. As we shall see, this
locus is the M67
WD cooling sequence and this is the first time that a clear WD cooling sequence
has been seen in this or in fact {\it any} old open cluster.

The CMD of the blank field reproduces the main features seen away from the
M67 cluster sequences but no very bright stars are seen as they were
saturated on our long exposures. The number of objects in the blank field CMD
is smaller than the total of field stars seen in the M67 frames as we only included in Figure 1
those stars which were present on the overlapping area of the blank 
field chips. Since these exposures were dithered by a large amount 
in order to construct a flat field, only about $70\%$ of the area covered 
by the M67
exposures is included in the right-hand panel of Figure 1. 
When we investigate the statistics of the objects found in both
the cluster and blank fields in the next section, we will have to account for
this by counting each observed blank field star $1.4$ times. In the blank 
field CMD there are a few
objects seen in the region occupied by the cluster WDs. From the data given
in \cite{fle86}, 11 field WDs ought to have been found
in an area the size covered by our cluster data brighter than $V = 23$ and 
bluer than $(V - I) = 0.3$. We actually count 8 such objects in the blank 
field which, coincidentally, scales up to 11 when we correct for the 
area difference. These 
will be removed statistically when we adjust the cluster counts for
background and foreground contamination.   

\section{The M67 White Dwarf Cooling Sequence}

\subsection{White Dwarfs in the Color-Magnitude Diagram}

Figure 2 replots the M67 CMD and superimposes on it cooling curves
for $0.7 M_{\odot}$ (upper) and $1.0 M_{\odot}$ pure carbon core DA (H-atmosphere) 
WDs. The cooling sequences were derived using the interior models of
\cite{woo95} and the model atmospheres of \cite{ber95} as discussed 
in \cite{ric95}, 1997. 
The $0.7 M_{\odot}$ cooling sequence is clearly a very
good fit to the WD data above an $M_V$ of $13.5$. This is likely to be
a reasonable estimate of the masses of M67 WDs it is the same as those
found in the Hyades (\cite{weg88}).
Fainter than $M_V = 13.5$ the
situation is confused by the presence of the large number of faint galaxies.
Even with this contamination there does appear to be a ``pile-up" of objects
in the CMD near $M_V = 14.7$ and $(V - I)_{\circ} = 0.7$. WDs will
tend to  ``pile-up" in a CMD as they cool more slowly as they age. Below, we 
will identify this
clump of objects with the termination of the WD cooling sequence in M67.
In the right-hand panel of Figure 2 we replot the $0.7 M_{\odot}$ WD cooling
sequence and indicate along it cooling times to different magnitudes. If the
background galaxies and the field stars can be successfully removed from the M67
CMD, the termination point of the WD cooling sequence will be located
and this can then yield a
lower limit to the age of the cluster. This age will only be a lower
limit since the main sequence lifetime of the progenitor of coolest WD 
must be added its cooling time to obtain the true cluster age.
However, if the coolest WDs seen in M67 evolved from massive main sequence
stars, the correction is minimal and the WD cooling time is an excellent
approximation of the cluster age. We also include in Figure 2 an isochrone
for an age of 4.0 Gyr and a metallicity of $[Fe/H] = -0.04$ (\cite{}hob91). This isochrone was kindly
supplied to us in advance of publication by \cite{vdb98} and is
clearly an excellent fit to the cluster CMD. We will use this age
below to compare with the WD cooling age of the cluster.

\subsection{The White Dwarf Cooling Age of M67}

We attempted to remove the background galaxies and field stars from the M67
CMD through the following procedure. 

(0) We determined incompleteness corrections in both the M67 and blank fields
by adding stellar PSFs of differing magnitudes directly
in to the frames and established the recovery statistics. The added
stars were placed along a $0.7 M_{\odot}$ WD cooling sequence shifted to M67's
distance 
and to be counted as a recovery the star had to be found in both the
$V$ and $I$ frames. The recovery statistics 
differed little between the blank and cluster fields so we combined them
together and found that down to $V = 24$ the completeness fraction was 1.0, 
from 
$24$ to $24.99$ it was $0.95$, from $25$ to $25.5$ it was $0.90$ and between
$25.5$ and $26.0$ it was $0.6$.

(1) We ran the image classifier 
SExtractor (\cite{ber96}) on all the objects
in the cluster and the blank field. SExtractor provides a class parameter which ranges 
from 0 to 1 with low numbers being galaxies and high numbers stars.
We then matched up the SExtractor objects with the output from  
ALLFRAME and plotted class versus the ALLFRAME sharp parameter
and $V$-magnitude for both the cluster and blank fields. 
These plots
indicated that a reasonable lower cut-off for the stars was 0.8. 
There are undoubtably some galaxies leaking in to the stellar
sample with little leakage in the other direction due to the excess
number of galaxies present compared to stars (to $V = 25$ in the cluster 
fields the 
galaxies outnumber the stars $3.6$ to $1$ while in the blank field the ratio is 
$8.7$ to $1$). We then removed all the 
objects classified as galaxies from our lists.

(2) We then rectified the $0.7 M_{\odot}$ cooling sequence to make it vertical
in a pseudo-CMD which plotted $V$ magnitude versus color distance from the
now vertical cooling sequence. The stars were included in this diagram
and we 
set up bins in both the cluster and blank field CMDs that were 
$0.5$ magnitudes wide in $V$ and $0.250$ wide on either side of the cooling sequence in
($V - I$). The width in color represents the $2-\sigma$ photometric errors at 
faint magnitudes. 
To correct for the smaller area in the blank field we multiplied the
number of objects in each bin in the blank field by $1.4$ and applied the incompleteness corrections to both data sets. 

(3) Lastly, we subtracted the blank field numbers from those in 
the cluster,
bin by bin, and calculated the errors associated with each bin. These
uncertainties 
include errors in the incompleteness corrections as well as the counting errors
but none from the Sextractor classification. 

In Figure 3 we plot, as a histogram, the results
of the above 
analysis. This diagram indicates a ``pile-up'' of WDs in the bin centered at
$V = 24$ and suggests that
the WD luminosity function terminates at
$V = 24.25$ ($M_V = 14.6$).
The cooling time
to this absolute magnitude for a $0.7 M_{\odot}$
pure carbon core WD with a hydrogen
atmosphere is 4.3 Gyrs. This age can
be compared with the isochrone age of the cluster of 4 Gyr shown in 
Figure 2. Changing the core composition to a mixture of carbon and
oxygen makes less than a $1\%$ difference in the age while changing the
atmosphere to He (DB WDs) has a similarly small effect. It is only for much
older WDs that such changes become important.

To assess the uncertainty in the above WD cooling age for M67, we note that 
had we taken the bin center as 
the termination point, the cooling age would be 3.5 Gyrs. 
If the masses of the cooling WDs were 
$0.1 M_{\odot}$ different than assumed, the cooling age would be 3.7 Gyr
($0.6 M_{\odot}$ WDs) or 4.5 Gyr ($0.8 M_{\odot}$ WDs).

\subsection{The White Dwarf Mass Fraction in M67}

As a last point we consider whether the number of WDs found in M67 is
consistent with expectations. The total number actually found is about 85 down
to the termination point of the WD cooling sequence and this
number must be corrected for those in binaries. As we demonstrate below, 
about 50\% of the M67 stars are in binary systems which likely means that
we must increase our estimate of the number of WDs by $50 - 100\%$ suggesting
a total population of about 150.  

From an analysis of the cluster dynamics, \cite{mat85} concluded that up
to half the mass of the cluster could be contained in WDs. If we take the 
total cluster
mass to be $800 M_{\odot}$ (\cite{fra89,mat85,mon93}) about
600 WDs ought to have been discovered, clearly at variance with our
results. 

We have rederived the cluster mass by simply adding 
up the mass of the cluster objects seen as we have observed the entire 
cluster with the exception of a few very bright giants which
were added in 
to our sample from the compilation of \cite{mon93}. Objects in various portions of the CMD were counted (after correcting
for the number of field stars) and assigned masses using the mass - $M_V$
relation of \cite{cha96} for faint main 
sequence stars and that 
of \cite{ber94} for $4$ Gyr for stars which
were above the Chabrier et al. limit but below the turnoff point. 
We chose as single
main sequence stars all those which fell along the fiducial of the main
sequence in a band $0.1$ magnitude wide in $(V-I)$. Binaries were chosen to be
all those stars above the upper limit of this band to a limit of $0.75$
magnitudes in $V$ above the middle of the main sequence. These objects had 
their masses 
doubled and were found to
contribute a substantial $44\%$ to the total cluster by mass. By number,
fully 50\% of the M67 stars are tied up in binary systems.
Stars 
above the turnoff were assigned a mass of $1.2 M_{\odot}$ and the WDs $0.7 M_{\odot}$. This yielded a mass of $1080 M_{\odot}$ with the
WDs contributing about $100 M_{\odot}$ of this, $9\%$ of the cluster total. 

The expected number of WDs can be estimated from complete model 
isochrones, such as those of \cite{cas92}.
To proceed, we adopt an 
initial mass function (IMF): $n({\cal M})\,\propto\,{\cal M}^{-(1+x)}$, with
$x = 1.35$. This (Salpeter) value 
is similar to that for field stars in the 
relevant mass range (\cite{kro93}), however,   
the results are insensitive to 
the precise value of $x$ for suitably short cooling times. The
mass function was normalized with the 87 post-turnoff stars 
counted in our CMD (this excludes blue and yellow stragglers) 
and the \cite{cas92} 4 Gyr
isochrone. This predicts $60 \pm 7$
WDs with
a cooling age $\le$ 1 Gyr. Similar numbers are obtained from less 
model-sensitive but  
cruder evolutionary timing arguments applied to the red giant clump stars. 
Since we observe only 24 WDs with cooling ages $\le$ 1 Gyr, 
this suggests that we are only seeing
$40 \pm 10$\% of the expected number of bright WDs.  
Some, or perhaps even the majority of the missing WDs are hidden in binaries 
(\cite{pas98,lnd97}). A further complication is the loss of stars through
dynamical evolution. M67 has a short relaxation time (\cite{mon93}) and an 
inverted mass function (\cite{fra89}), suggesting a significant loss
of stars. The large binary population in M67 not only affects 
the morphology of the cluster CMD, complicating the interpretation of the 
luminosity function, but also strongly influences the dynamical evolution. 
A better understanding of these effects is required to fully interpret  
the WD luminosity function in M67.

%
%

\clearpage

\figcaption[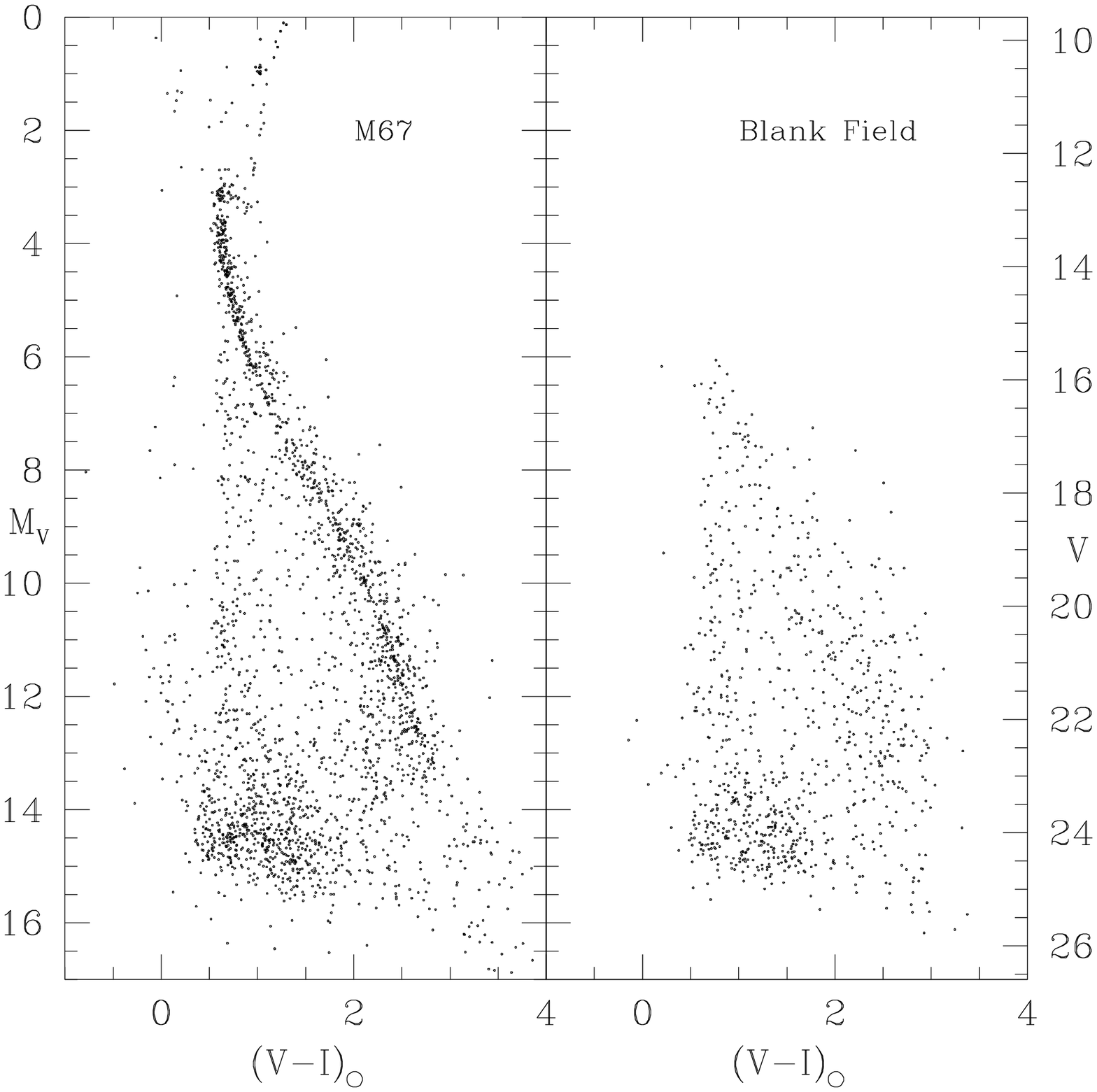]{In the left panel the M67 $M_V$, $(V - I)_{\circ}$ CMD
for the {\it entire} cluster is shown. On the right is the CMD for 
the blank field. Only objects passing a shape test indicating that they were
likely to be stars are included in these diagrams.  \label{fig1}}

\figcaption[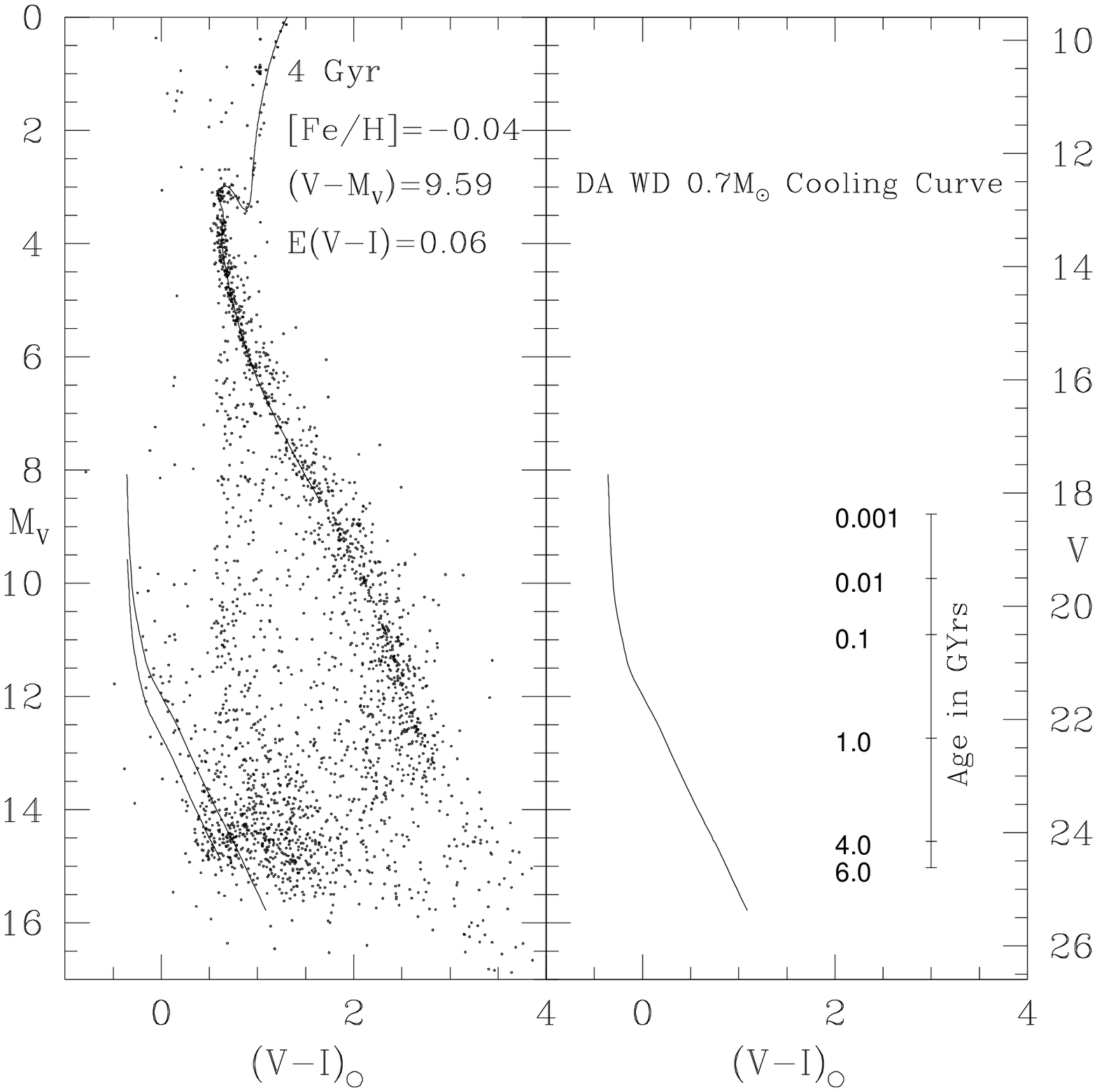]{The left panel here is the same as in Figure 1 except that
theoretical cooling curves for $0.7$ (upper) and $1.0 M_{\odot}$ DA WDs are
included. Also shown is an isochrone for 4 Gyr for a metallicity 
of $[Fe/H] = -0.04$ which is appropriate to M67. In the right panel we 
replot the $0.7 M_{\odot}$
cooling curve and indicate along it cooling times to various magnitudes.
\label{fig2}}

\figcaption[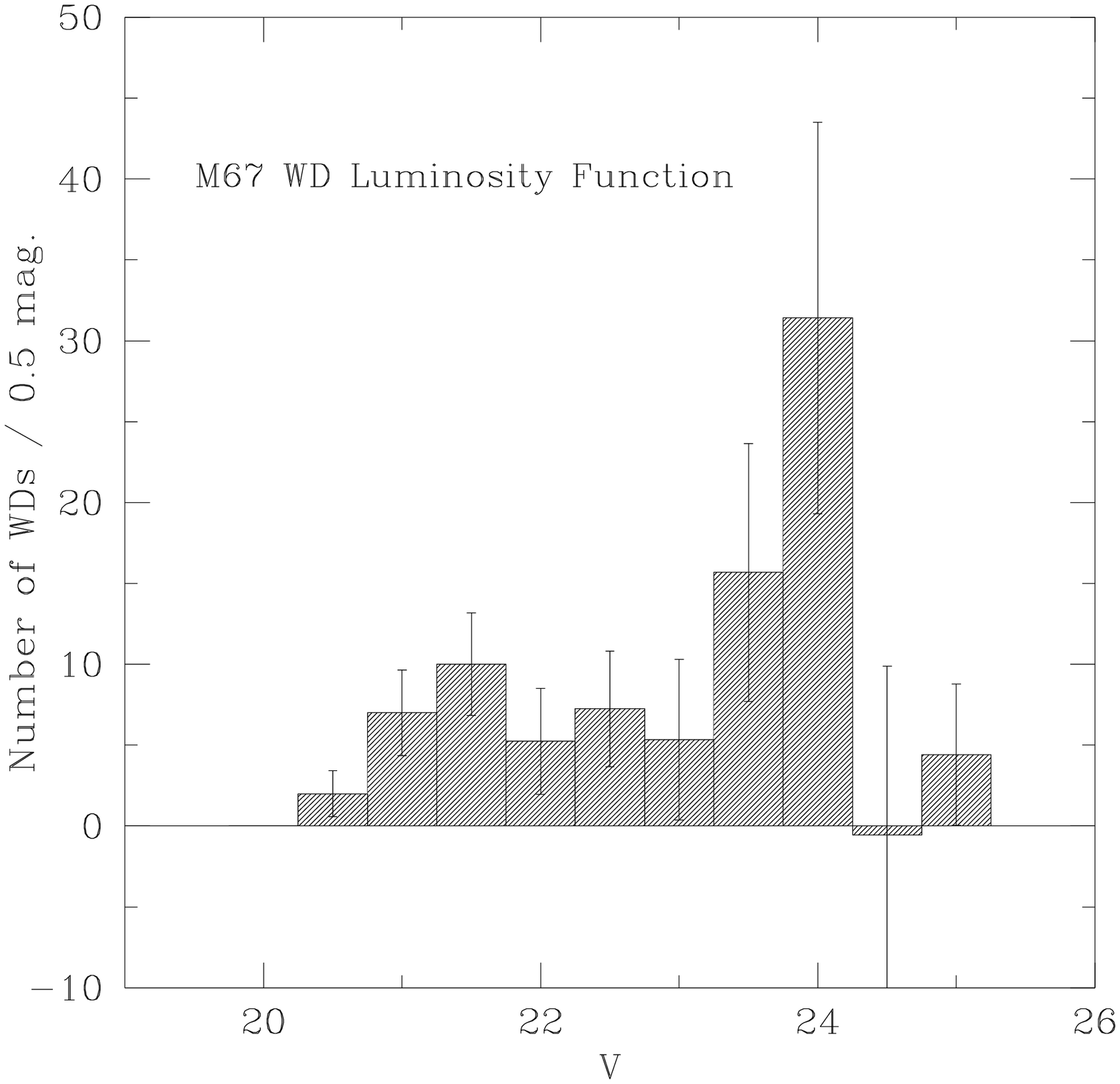]{The M67 WD luminosity function. There appears to be
a termination of the luminosity function at $V = 24.25$ ($M_V = 14.6$) which
corresponds to a cooling time of 4.3 Gyr for $0.7 M_{\odot}$ WDs.
 \label{fig3}}

\end{document}